\documentclass{aa}  

\usepackage{natbib}
\bibpunct{(}{)}{;}{a}{}{,}

\usepackage{graphicx}
\usepackage{txfonts}
\usepackage{hyperref}
\usepackage{lipsum}
\usepackage{booktabs}
\usepackage{mathrsfs}
\DeclareMathAlphabet{\mathcal}{OMS}{cmsy}{m}{n}
\usepackage[shortlabels]{enumitem}
\usepackage{subcaption}
\usepackage{lscape}
\usepackage{placeins}

\begin{document}

   \title{Deep-learning-based prediction of precipitable water vapor in the Chajnantor area}

   \author{Alison Matus-Bello\inst{1,2,3,4}
        \and Silvia E. Restrepo\inst{5,6}
        \and Ricardo Bustos\inst{5}
        \and Yi Hu\inst{7}
        \and Fujia Du\inst{8,9}
        \and Jaime Cariñe\inst{5}
        \and Pablo García\inst{10,11}
        \and Javier Maldonado\inst{4}
        \and Rodrigo Reeves\inst{12}
        \and Zhaohui Shang\inst{7}
        }

\authorrunning{A. Matus-Bello et al.}
\titlerunning{Deep learning-based prediction of PWV in the Chajnantor area}

   \institute{Facultad de Ingeniería, Universidad Católica de la Santísima Concepción, Alonso de Ribera 2850, Concepción, Chile\\
             \email{amatus@doctoradoia.cl}
             \and Departamento Ingeniería Informática y Ciencias de la Computación, Universidad de Concepción, Concepción 4070409, Chile
             \and Facultad de Ingeniería, Universidad del Bío-Bío, Collao 1202, Concepción, Chile
             \and Departamento de Electrónica e Informática, Universidad Técnica Federico Santa María - Sede Concepción, Arteaga Alemparte 943, Concepción, Chile
             \and Departamento de Ingeniería Eléctrica, Universidad Católica de la Santísima Concepción, Alonso de Ribera 2850, Concepción, Chile
             \and Centro de Energía, Universidad Católica de la Santísima Concepción, Alonso de Ribera 2850, Concepción, Chile
             \and National Astronomical Observatories, Chinese Academy of Sciences, Beijing, 100101, China
             \and Nanjing Institute of Astronomical Optics \& Technology, Chinese Academy of Science, Nanjing, 210042, China
             \and CAS Key Laboratory of Astronomical Optics \& Technology, Nanjing Institute of Astronomical Optics \& Technology, Nanjing, 210042, China
             \and Chinese Academy of Sciences South America Center for Astronomy, National Astronomical Observatories, CAS, Beijing 100101, China
             \and Instituto de Astronomía, Universidad Católica del Norte, Av. Angamos 0610, Antofagasta, Chile
             \and CePIA, Departamento de Astronomía, Universidad de Concepción, Casilla 160 C, Concepción, Chile}
            
   \date{Received XXXX; accepted XXXX}

  \abstract
   {Astronomical observations at millimeter and submillimeter wavelengths strongly depend on the amount of precipitable water vapor (PWV) in the atmosphere, which directly affects the sky transparency and decreases the signal-to-noise ratio of the signals received by radio telescopes.}
   {Predictions of PWV at different forecasting horizons are crucial to supporting the telescope operations, engineering planning, scheduling and observational efficiency of radio observatories installed in the Chajnantor area in northern Chile.} 
   {We developed and validated a long short-term memory (LSTM) deep-learning-based model to predict PWV at forecasting horizons of 12, 24, 36, and 48 hours using historical data from two 183 GHz radiometers and a weather station in the Chajnantor area.}
   {We find that the LSTM method is able to predict PWV in the 12- and 24-hour forecasting horizons with a mean absolute percentage error of \string~22\% compared to \string~36\% for the traditional Global Forecast System method used by the Atacama Pathfinder Experiment, and the root mean square error is reduced by \string~50\%.}
   {We present a first application of deep learning techniques for preliminary predictions of PWV in the Chajnantor area. The prediction performance shows significant improvements compared to traditional methods in 12- and 24-hour time windows. We also propose strategies to improve our method on shorter (< 12 hour) and longer (> 36 hour) forecasting timescales.}


   \maketitle

\section{Introduction}
Astronomical observations at millimeter and submillimeter wavelengths heavily depend on atmospheric conditions, particularly the amount of precipitable water vapor (PWV). This parameter describes the thickness of the water column, measured in millimeters, that would result from collapsing all the water vapor contained in the atmospheric column along the observer's line of sight at a specific geographical location. The presence of PWV directly affects sky transparency, decreasing the signal-to-noise ratio of the signals received by radio telescopes and increasing the integration time required to detect faint astronomical sources \citep{radford_observing_2011}. High levels of PWV can render scientific observations in certain bands impossible. 

The Chajnantor area, located in the Atacama Desert in northern Chile, is recognized as one of the best sites in the world for (sub)millimeter astronomical observations due to its high altitude (\string~5000 m) and extreme atmospheric dryness \citep{bustos_parque_2014,radford_submillimeter_2016}. This location hosts cutting-edge observatories such as Atacama Large Millimeter/submillimeter Array (ALMA), Atacama Pathfinder EXperiment (APEX), Cosmology Large Angular Scale Surveyor (CLASS), Simons Observatory, and many others. Predicting PWV enables engineers and astronomers to optimize operations and the use of valuable telescope time by scheduling their observing frequencies for times with the most favorable atmospheric conditions, thus maximizing scientific return \citep{cortes_twenty_2020}. The highly variable nature of PWV at these altitudes requires reliable predictive tools. In this context, accurate monitoring and forecasting of PWV is crucial to supporting telescope operations, engineering planning, and observational scheduling. 

Traditionally, PWV has been estimated using, for example, radiometers operating at specific water vapor emission lines (e.g., 183 GHz), radiosondes, and global navigation satellite systems (GNSSs). Radiometers can provide continuous and direct measurements of PWV, including at remote high-altitude locations with extreme environmental conditions \citep{otarola_thirty_2010,otarola_precipitable_2019,valeria_satellite-based_2024}. Using GNSSs, properly calibrated, is a powerful technique for obtaining PWV by converting zenith tropospheric delays into water vapor content measurements with high accuracy \citep{wu_high-precision_2021,chen_global_2021}. \citet{castro-almazan_precipitable_2016} calibrated GNSS-based PWV estimates with radiosonde profiles to obtain the PWV at observatories in the Canary Islands. Recent advances in weighted mean temperature modeling have further improved GNSS retrievals as demonstrated by \citet{sleem_new_2024}, who developed a high-resolution empirical model based on ERA5 data, a state-of-the-art atmospheric reanalysis produced by the European Centre for Medium-Range Weather Forecasts (ECMWF). Also, \citet{li_research_2024} show that real-time single-station GNSSs can deliver reliable PWV estimates extending applications to coastal and offshore regions. A comprehensive review by \citet{vaquero-martinez_review_2021} summarizes three decades of GNSS meteorology, emphasizing its crucial role in PWV monitoring, numerical model validation, and climate analysis. While these methods provide valuable real-time measurements, they primarily address PWV monitoring rather than proactive forecasting.

Forecasting PWV often relies on numerical weather prediction (NWP) models, such as the Global Forecast System (GFS). While NWP models provide broad geographical coverage, they typically lack the fine-grained spatial and temporal resolution needed to accurately capture the rapid, localized variations in atmospheric moisture at high-altitude observatory sites. This limitation significantly impacts the accuracy of short-term forecasts, which are critical for scheduling. PWV forecasts by \citet{marin_estimating_2015} in the Chajnantor area based on Geostationary Operational Environmental Satellites and GFS data had errors of 33\% in the 0.4–1.2\,mm range compared with PWV observations. The GFS model is currently being applied at the APEX telescope \citep{gusten_atacama_2006} and provides 5-day forecasts in 6-hour step resolution. 

Previous studies have rigorously evaluated  approaches both on GNSSs and weather research and forecasting (WRF) for PWV estimation and forecasting at major astronomical observatories. For instance, \citet{perez-jordan_precipitable_2018} validated the WRF model for PWV forecasting at the Roque de los Muchachos Observatory. Similarly, \citet{pozo_validation_2016} assessed the performance of the WRF model at the Chajnantor Plateau, finding that forecast errors are greater than 1.5\,mm in 10\% of cases and do not exceed 0.5\,mm 65\% of the time.

Similar forecasting approaches using NWP models have also been successfully implemented at other major astronomical sites. For example, at the Very Large Telescope at Cerro Paranal and the Large Binocular Telescope at Mount Graham, using the Meso-NH mesoscale atmospheric model, has improved the European Centre for Medium-Range Weather Forecasts by a factor of 2 \citep{turchi_forecasting_2018}. Likewise, the WRF model has been validated at Observatorio del Roque de los Muchachos \citep{giordano_atmospheric_2013} and at APEX telescope \citep{pozo_validation_2016}. All this demonstrates the widespread adoption of NWP tools in the astronomical community for PWV predictions.

Global navigation satellite system data have also been increasingly assimilated into NWP systems. For example, \citet{li_investigating_2023} demonstrate that assimilating GNSS PWV measurements into the WRF model improved humidity field representation by up to 26\% in Australia during heavy rainfall events. Similarly, \citet{wu_high-precision_2021} report that GNSS assimilation across China significantly enhanced long-term PWV climatology and precipitation forecasts. These results underline the dual role of GNSSs both as an independent observation system and as a source of assimilation data for NWP.

In response to the inherent limitations of traditional techniques and the insufficient resolution of NWP models for site-specific forecasting, alternative and advanced data-driven approaches have emerged as powerful predictive tools. In particular, deep learning models have shown the ability to capture nonlinear dependences and temporal patterns in complex atmospheric datasets. Among these, long short-term memory (LSTM) networks stand out for their capacity to handle sequential data and forecast variables with high accuracy. Studies of recent applications in atmospheric science report that such models often outperform traditional statistical and physical methods when trained with local, high-resolution data \citep{xiao_prediction_2022,hou_machine_2023,yan_research_2024}.

Together, these advances highlight the current diversity of methods: GNSSs provide accurate retrievals and assimilation capacity, NWP models such as WRF offer physically based forecasts, and deep learning approaches add the potential to capture nonlinear dependences in site-specific high-frequency datasets. The present study contributes by applying a deep learning framework to continuous 183\,GHz radiometer data combined with meteorological records.

In this context, the objective of the present study is to develop and validate a preliminary deep-learning-based model to predict PWV on 12-, 24-, 36- and 48-hours forecasting horizons using one year of 183 GHz radiometer and meteorological data for the Chajnantor area. In Section 2, we present the data and methodology. Section 3 describes the LSTM model. Section 4 presents the results obtained and strategies to improve our method and Section 5 presents our conclusions. 

\section{Data and methodology}
The dataset employed in this study spans from July 13, 2023, to October 14, 2024 (460 days), covering one full annual cycle at the Chajnantor Plateau. It integrates measurements from two 183 GHz radiometers and a meteorological station.

\subsection{183 GHz radiometer data}
We used PWV data collected by two 183\,GHz radiometers. One radiometer is located at the APEX telescope in Llano de Chajnantor at 23º 00' 20.8"\,S, 67º 45' 33.0"\,W and an altitude of 5,105\,masl. The second radiometer is operated by the Universidad Católica de la Santísima Concepción (UCSC) in collaboration with the Universidad de Concepción (UdeC) and is located at the CLASS telescope on Cerro Toco at 22º 57' 34.6"\,S, 67º 47' 13.8"\,W and 5,200\,masl. The horizontal distance between the two sites is approximately 5.9\,km.

Both instruments provide continuous data. PWV from APEX (PWV APEX) is recorded as 1-minute averages, while the UCSC radiometer (PWV UCSC) samples every 2 seconds and is averaged to 1-minute intervals, yielding approximately 662,400 records per variable. This configuration allows redundancy and cross-validation of the measurements, given the small altitude difference between the two sites (\string~100,m).
To evaluate our model at different atmospheric conditions, PWV values were stratified into three categories: $<$1\,mm (33.4\% of the records), 1–2\,mm (31.1\%), and $>$2\,mm (35.5\%). We also studied the PWV range 0.4-1.2\,mm to compare with current GFS predictions.

\subsection{Meteorological data}
The meteorological data were obtained from the automated weather station at the APEX telescope site. This station provides continuous monitoring of local atmospheric conditions such as relative humidity (RH) in percentage, temperature (Temp) in degrees Celsius, wind speed (WS) in meters per second, and wind direction (WD) in degrees. The raw data were collected at one-minute resolution, synchronized with the radiometric series.  

These observations complement radiometer measurements by providing atmospheric context. Together, they enable an integrated analysis of PWV variability with local meteorological conditions, essential for both operational forecasting and scientific applications. Meteorological data, PWV measurements, and GFS forecast data from APEX are publicly available \footnote{\url{http://www.apex-telescope.org}}.

\subsection{Data preprocessing}
To ensure the reliability of the predictive model, a preprocessing workflow was applied to the raw dataset. The workflow involved several sequential steps applied to the initial 1-minute resolution data: 1) data cleaning and outlier removal; 2) linear interpolation to fill missing values; 3) temporal averaging to 3-hour intervals; 4) dataset selection of meteorological variables; 5) Fourier analysis; 6) wavelet-based denoising filter; and 7) z-score normalization of all features. 

\subsubsection{Data cleaning}
Outliers were identified using the interquartile range method, which detects anomalous points lying beyond 1.5 times the interquartile distance below Q1 or above Q3. Instrumental data flagged as $\pm$999 were also removed before interpolation with time-aware linear methods.

\subsubsection{Imputation of missing values}
The dataset used in this study, composed of PWV UCSC, PWV APEX, Temp, RH, WS, and WD, contained missing data in several variables. PWV UCSC had 0.7\% of missing values, while PWV APEX had 20.7\%. Temp, RH, WS, and WD each had 8.1\% missing entries. These gaps were distributed throughout the year and likely resulted from instrumental issues, such as sensor saturation, temporary outages, or data logging interruptions.

To ensure continuity in the time series, missing values were imputed using linear interpolation. This method was selected for its simplicity and its effectiveness in estimating short, isolated gaps (typically under 30 minutes) without introducing artificial trends or discontinuities.

Since the data were subsequently averaged into 3-hour intervals, the interpolation had minimal influence on the overall signal structure. This approach helped to preserve temporal consistency across variables and ensured that the LSTM model received complete, synchronized input sequences.

\subsubsection{Correlation and input parameter selection}
A Pearson correlation analysis was performed to guide the variable selection process, retaining only those with a statistically significant relationship to the target variable, PWV APEX. The Pearson correlation coefficient ($r$) is defined as

\begin{equation}
r = \frac{\sum_{i=1}^{n} \left[(x_i - \bar{x})(y_i - \bar{y})\right]}
{\sqrt{\sum_{i=1}^{n}(x_i - \bar{x})^2 \sum_{i=1}^{n}(y_i - \bar{y})^2}}.
\label{eq:pearson}
\end{equation}

Where $x_i$ and $y_i$ are the samples of each variable, $\bar{x}$ is the mean of the $x$ values, $\bar{y}$ is the mean of the $y$ values, and $n$ is the total number of observations. PWV APEX corresponds to $y$.

\begin{figure}[h!]
\centering
\includegraphics[width=\hsize]{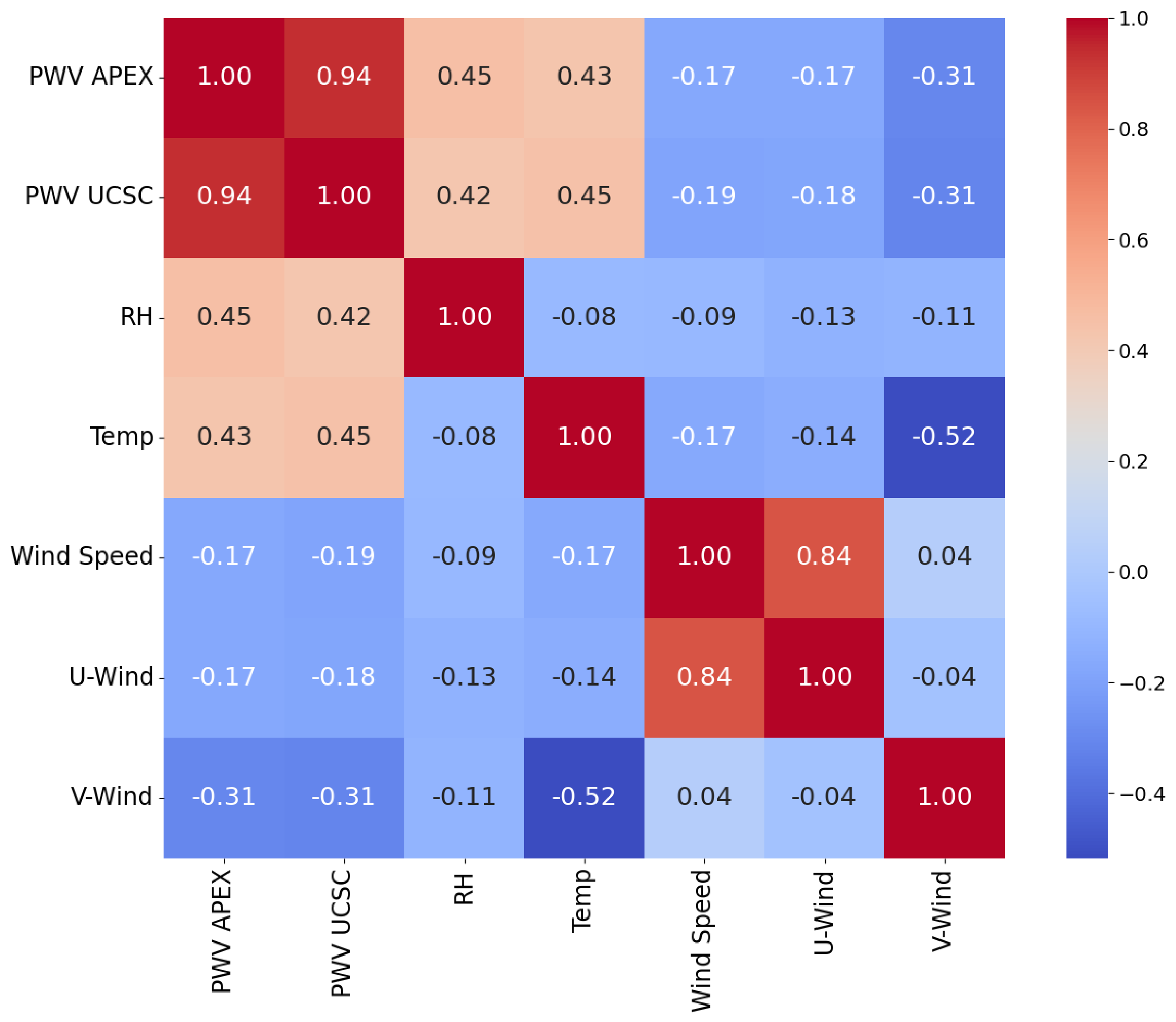}
\caption{Correlation matrix for PWV APEX, PWV UCSC, RH, Temp, WS, and $U$- and $V$-wind components. Warmer colors are positive correlations, while cooler colors are negative correlations.}
\label{correlation}
\end{figure}

The analysis seen in Figure~\ref{correlation} revealed an expected strong positive correlation between PWV APEX and PWV UCSC ($r = 0.94$), and moderate correlations with RH ($r = 0.45$) and Temp ($r = 0.43$). In contrast, WS and the $U$ and $V$ wind components showed negligible correlations (all $|r| < 0.20$) and were excluded from the model due to their limited predictive value. As a result, the selected input variables are: (1) PWV APEX, (2) PWV UCSC, (3) Temp, and (4) RH.

\FloatBarrier 

\subsubsection{Temporal averaging}
To reduce sensitivity to high-frequency noise and short-term variability, all selected time series were averaged into 3-hour intervals, a resolution that facilitates the identification of broader atmospheric trends (12–48 h horizons) and aligns with operational forecasting models such as GFS. These four variables were resampled and synchronized on the same temporal basis.

After the previous steps of data preprocessing, the data are resampled to 3-hour temporal resolution. For the radiometric data, PWV APEX ranged from 0.14 to 8.21\,mm, with a mean of 2.00\,mm and a median of 1.46\,mm, while PWV UCSC ranged from 0.12 to 6.03\,mm, with a mean of 1.72\,mm and a median of 1.33\,mm. For the meteorological data, RH ranged from 0.26\% to 100.2\%, with a mean of 30.2\% and a median of 24.3\%, while Temp ranged from -17.05\,°C to 10.55\,°C, with a mean of -1.65\,°C and a median of –1.38\,°C.

\begin{figure}[h!]
   \centering
   \includegraphics[width=\hsize]
   {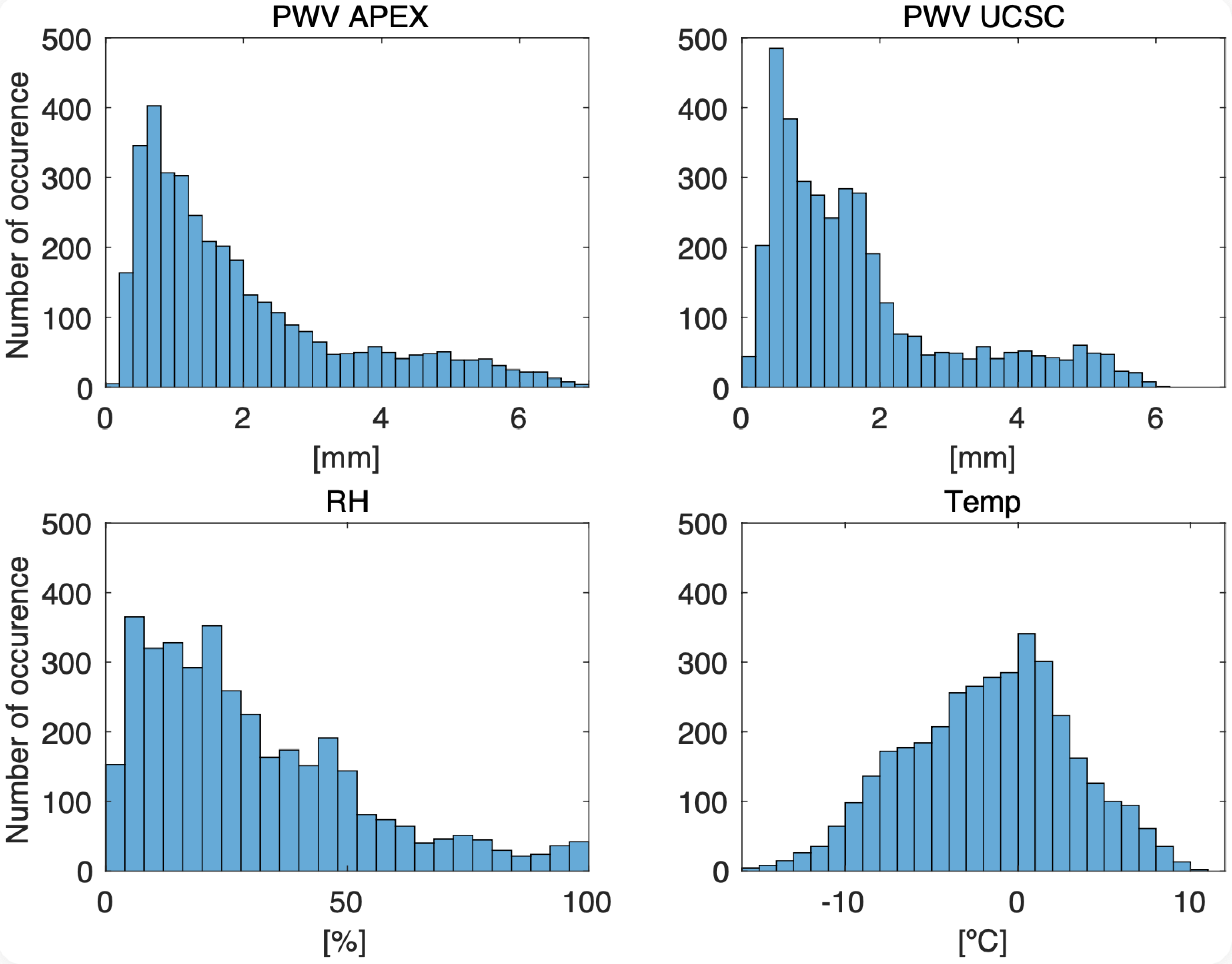}
      \caption{Number of occurrences for each variable after data cleaning, imputation of missing values, and 3-hour averaging.}
         \label{hist}
   \end{figure}

The distribution of the values is shown in Figure~\ref{hist}. The data series contains a total of 3,672 records per variable and the binning for PWV is 0.2\,mm, for RH is 4\%, and for Temp is 1\,°C.

\subsubsection{Fourier analysis}
\begin{figure}[h!]
\centering
\includegraphics[width=\hsize]{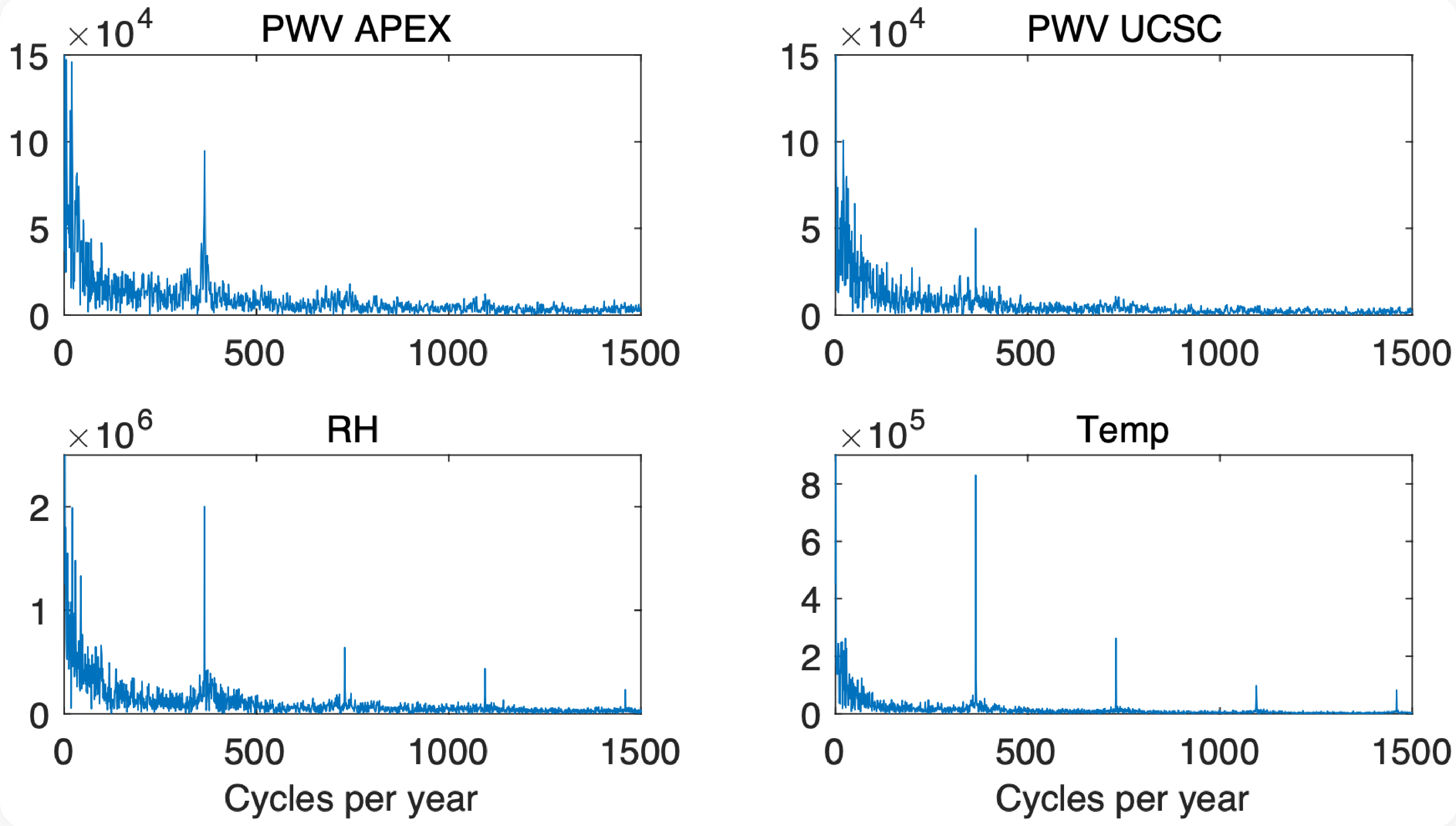}
\caption{Power spectral density plots from the FFT analysis of PWV APEX, PWV UCSC, RH, and Temp. The x-axis represents frequency in cycles per year.}
\label{fft}
\end{figure}

To better understand the dominant periodicities and their cyclic behavior in PWV and meteorological variables, a fast Fourier transform (FFT) was applied. The FFT obtained for each variable is seen in Figure~\ref{fft}, revealing clear peaks at frequencies corresponding to a yearly cycle (1 per year), daily cycles (365 cycles per year), and sub-daily harmonics, indicating that these signals contain predictable recurring patterns. These insights supported the decision to use sequential models such as LSTM, which are capable of capturing temporal dependences.

\FloatBarrier 

\subsubsection{Wavelet denoising filter}
To improve the signal-to-noise ratio, a wavelet-based denoising technique was applied using the Symlet 4 (Sym4) wavelet function. This wavelet was selected after testing alternatives such as Daubechies 4 (db4), Coiflet 4 (coif4), and Haar, offering the best balance between noise reduction and preservation of key signal trends.

In this context, noise in the PWV data primarily stems from instrumental fluctuations and abrupt environmental changes, which introduce high-frequency variations that can obscure relevant temporal patterns. These distortions reduce the model's ability to learn consistent relationships.

The wavelet filter effectively suppresses high-frequency noise while retaining low-frequency components crucial for trend analysis. As a result, the smoothed signals exhibit reduced short-term variability and improve model training stability and accuracy.

\subsubsection{Data normalization}
All selected input variables were normalized using z-score standardization (zero mean and unit variance). This step ensures that differences in magnitude among variables do not bias the model training process and allows the LSTM network to converge more efficiently. With all variables normalized and the input structure defined, the four selected datasets at 3-hour intervals were prepared for training and evaluation using the deep learning model.

\section{Deep learning model: LSTM}
In this study, a deep learning model based on recurrent neural networks, specifically long short-term memory (LSTM), was developed for the prediction of PWV. These networks are particularly suitable for sequential data analysis due to their ability to handle both short- and long-term temporal relationships. This is made possible by their memory cell-based architecture, which surpasses the limitations of traditional recurrent networks, such as gradient vanishing or explosion \citep{hochreiter_long_1997,hou_machine_2023,haputhanthri_short-term_2021,jain_forecasting_2020}.

The model uses as input the selected predictors (PWV UCSC, Temp, RH), together with historical values of the target variable (PWV APEX), to forecast future PWV APEX values. To ensure the integrity of the predictive process and avoid biases known as ``data leakage'', only past values were used as inputs and targets, replicating realistic forecasting conditions.

The network was configured to analyze temporal windows of 48 steps, equivalent to 6 days of data. The architecture includes two hidden layers of 120 neurons each. This configuration was selected after conducting multiple tests, where different hidden layer sizes were compared. The final setup was chosen based on the best performance in terms of lowest root mean square error (RMSE) and highest coefficient of determination ($R^{2}$) across the 12 and 24\,h forecasting horizons.

The Adam optimizer was employed for its ability to dynamically adjust the learning rate, which improves model convergence \citep{yang_adaptability_2022}. This provided the best balance between convergence speed and performance compared to other optimizers such as AdamW, stochastic gradient descent, and RMSprop. The mean squared error was employed as the loss function.

\begin{figure}[h!]
\centering
\includegraphics[width=\hsize]{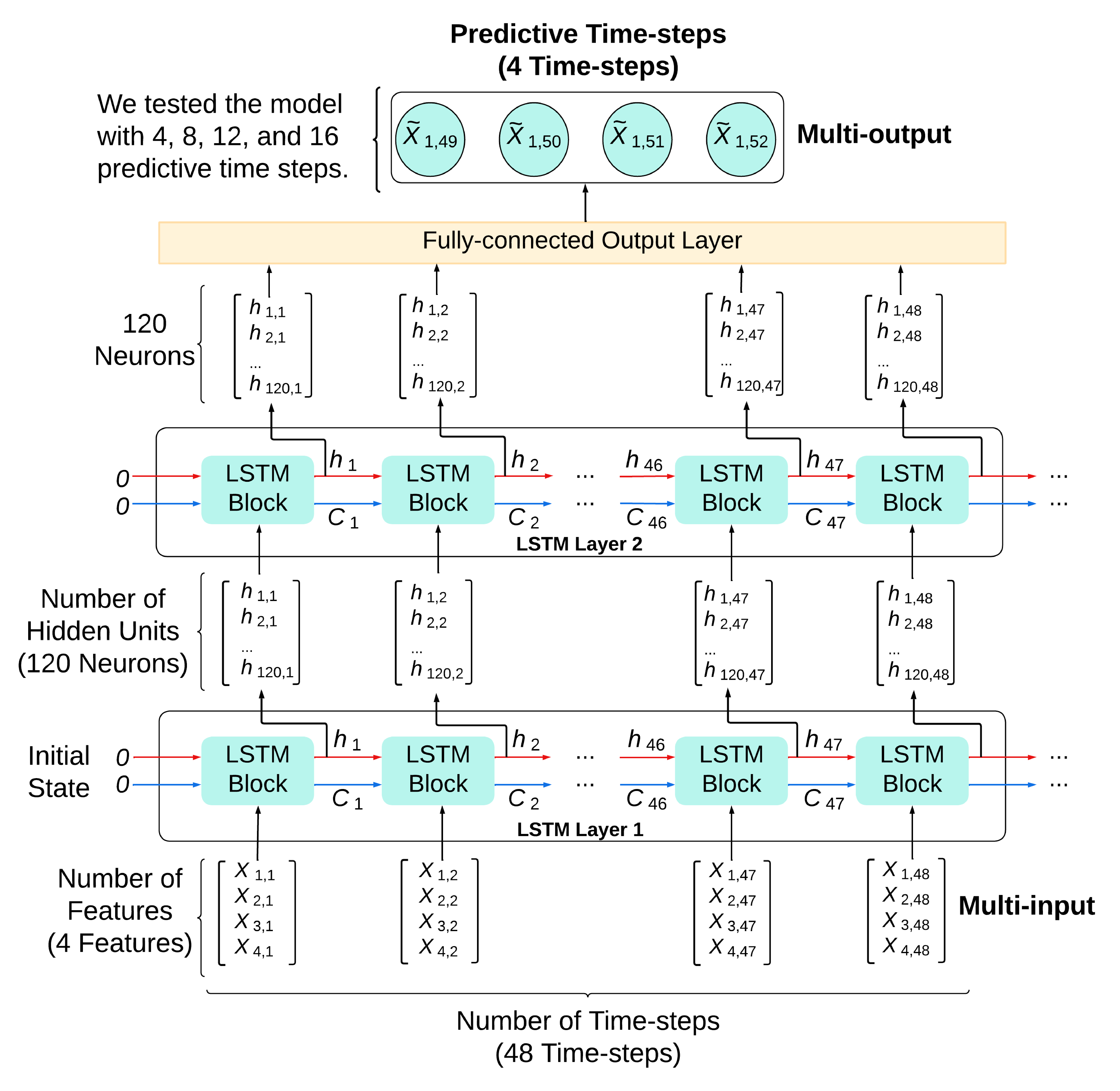}
\caption{Diagram of the LSTM architecture.}
\label{lstm_arch}
\end{figure}

Figure~\ref{lstm_arch} presents the architecture of the proposed LSTM model. The model takes as input four variables (Features): PWV APEX (1), PWV UCSC (2), Temp (3), and RH (4) recorded over 48 time-steps. The output consists of predicted values of future PWV APEX at four distinct time horizons: 4, 8, 12, and 16 time-steps ahead, which correspond to 12, 24, 36, and 48 hours into the future, respectively (multi-output).

\FloatBarrier

The architecture includes two stacked LSTM layers, each with 120 hidden units (neurons), which process the input sequence to extract relevant temporal patterns. These are followed by a fully connected output layer that maps the final LSTM outputs to the predicted values. This configuration was selected for its ability to effectively capture complex temporal dynamics while maintaining efficient training performance.

\subsection{LSTM architecture}
The LSTM network is a type of recurrent neural network designed to address the limitations of standard recurrent neural networks when modeling long sequences. It incorporates a memory cell regulated by three gates (input, forget, and output) that control the flow of information over time. These gates determine which information is retained, discarded, or passed to the next time step.

The following formulation is adapted from standard LSTM definitions, using the approach by \citet{hou_machine_2023}. At each time step \( t \), the gates are calculated as follows:

\begin{align}
i_t &= \sigma(W_i \cdot [h_{t-1}, x_t] + b_i) \label{eq:input_gate} \\
f_t &= \sigma(W_f \cdot [h_{t-1}, x_t] + b_f) \label{eq:forget_gate} \\
o_t &= \sigma(W_o \cdot [h_{t-1}, x_t] + b_o) \label{eq:output_gate}
\end{align}

In these expressions, \( x_t \) is the input vector at time step \( t \), and \( h_{t-1} \) is the hidden state from the previous time step. The weight matrices \( W_i, W_f, W_o \) and the bias vectors \( b_i, b_f, b_o \) are learned during training. The sigmoid activation function \( \sigma \) constrains the gate outputs to values between 0 and 1, effectively controlling how much information passes through each gate. The candidate cell state \( \tilde{C}_t \), which represents new information to be considered for memory update, is computed as:

\begin{equation}
\tilde{C}_t = \tanh(W_c \cdot [h_{t-1}, x_t] + b_c) \label{eq:candidate_state}
\end{equation}

Here, \( W_c \) and \( b_c \) are the weight matrix and bias associated with the candidate state, and \( \tanh \) is the hyperbolic tangent activation function. The cell state \( C_t \) is updated by combining the previous state with the candidate state, modulated by the forget and input gates, respectively:

\begin{equation}
C_t = f_t \odot C_{t-1} + i_t \odot \tilde{C}_t \label{eq:cell_state}
\end{equation}

The symbol \( \odot \) denotes the Hadamard (element-wise) product, allowing for selective control over each element of the state vector. Finally, the hidden state \( h_t \), which also serves as the output of the LSTM cell, is computed using the output gate and the updated cell state:

\begin{equation}
h_t = o_t \odot \tanh(C_t) \label{eq:hidden_state}
\end{equation}

This mechanism enables the LSTM network to maintain long-term dependences in sequential data, while adaptively filtering relevant information at each time step.

\subsection{LSTM model implementation}
The model was implemented in MATLAB R2024b and Python 3.8.20 for cross-validation. The dataset was split into training (80\%) and validation (20\%). The training data covered the period from July 13, 2023, to July 14, 2024, while the validation data ranged from July 15, 2024, to October 14, 2024. Training was conducted over 180 epochs using the Adam optimizer.

\subsection{LSTM model evaluation}
The model's performance was evaluated using multiple metrics, including the correlation coefficient ($r$), coefficient of determination ($R^2$), RMSE, mean absolute error (MAE), and mean absolute percentage error (MAPE). 

The $R^2$ measures the proportion of variability in the actual data explained by the model. Values close to 1 indicate an excellent fit. $R^2$ is defined as:

\begin{equation} 
\label{eq:r2} R^2 = 1 - \frac{\sum_{i=1}^{n} (\hat{y}_i - \bar{y})^2}{\sum_{i=1}^{n} (y_i - \bar{y})^2}
\end{equation}

where $y_i$ denotes the observed (true) values, $\hat{y}_i$ is the predicted value for each $i$, $\bar{y}$ is the mean of the observed values, and $n$ is the total number of observations.

The RMSE was also employed to measure the average magnitude of errors between predictions and actual values, giving greater weight to larger errors. Its interpretation is intuitive, as it is expressed in the same units as the target variable (i.e., in millimeters).

\begin{equation}
\label{eq:rmse}
\mathrm{RMSE} = \sqrt{\frac{1}{n} \sum\limits_{i=1}^{n} (y_i - \hat{y}_i)^2}
\end{equation}

Additionally, the MAE was used as a complementary metric, representing the average absolute difference between predicted and actual values. Unlike RMSE, MAE treats all errors equally without emphasizing larger deviations, making it useful for assessing overall prediction performance.

\begin{equation} 
\label{eq:mae} \mathrm{MAE} = \frac{1}{n} \sum_{i=1}^{n} |y_i - \hat{y}_i| 
\end{equation}

Finally, the MAPE was calculated to express the average error as a percentage relative to the actual values. This metric is particularly useful for evaluating model performance across different scales, although it may be less stable when actual values approach zero. To further assess the model’s performance, we generated a set of plots, including the evolution of loss during training, predicted versus actual value comparisons, and scatter plots with regression lines.

\begin{equation} 
\label{eq:mape} \mathrm{MAPE} = \frac{1}{n} \sum_{i=1}^{n} \left| \frac{y_i - \hat{y}_i}{y_i} \right| \times 100\% 
\end{equation}

\section{Results and discussion}

This section presents the results obtained from LSTM predictions of PWV APEX compared with real future values of PWV measured at APEX. The performance was evaluated for different forecasting of 4, 8, 12 and 16 steps, corresponding to forecasting horizons of 12, 24, 36, and 48 hours.

\subsection{Comparative evaluation of different forecasting horizons}

Table \ref{table:lstm_performance} presents the performance of the LSTM model across the four forecasting horizons. The results show a strong correlation $r$ and high $R^2$ for short-term predictions (12 and 24\,h) of real PWV values. In these two forecasting horizons, RMSE is \string~0.35\,mm, MAE \string~0.24\,mm, and MAPE \string~22\%.

\begin{table*}[!t]
\caption{Performance of the LSTM model for different forecasting horizons}
\label{table:lstm_performance}
\centering
\begin{tabular}{ccccccc}
\hline\hline
Steps & Time & $r$ & $R^2$ & RMSE (mm) & MAE (mm) & MAPE (\%)\\
\hline
4  & 12 h & 0.9492 & 0.9009 & 0.328  & 0.230 & 21.1 \\
8  & 24 h & 0.9336 & 0.8704 & 0.376  & 0.256 & 23.2 \\
12 & 36 h & 0.6917 & 0.4583 & 0.768  & 0.547 & 62.8 \\
16 & 48 h & 0.5578 & 0.2297 & 0.916  & 0.614 & 65.9 \\
\hline
\end{tabular}
\end{table*}

The prediction accuracy decreases as the forecasting horizon increases to 36 and 48\,h. Quantitatively, the correlation coefficient decreases from $r=0.95$ (12 h) and $r=0.93$ (24 h) to $r=0.69$ (36 h) and $r=0.56$ (48 h), while $R^2$ drops from 0.90 and 0.87 to 0.46 and 0.23, respectively. RMSE values increase from 0.33 mm (12\,h) and 0.38\,mm (24\,h) to 0.77\,mm (36\,h) and 0.92\,mm (48\,h), and MAPE increases dramatically from 21.1\% (12\,h) and 23.2\% (24\,h) to over 60\% (36 and 48\,h). These indicators highlight that both absolute and relative errors grow rapidly once forecasts extend beyond one day. An RMSE greater than 0.9\,mm at 48\,h implies a considerably large uncertainty in atmospheric conditions, potentially impacting the feasibility or quality of these predictions.

This indicates that LSTM model can reliably predict a significant portion of the observed PWV in time frames of 12 and 24 h, suggesting the utility of these predictions for operational tasks at observatories sensitive to PWV conditions.

\subsection{Comparison between predicted and real PWVs}

\begin{figure}[h!]
   \centering
   \includegraphics[width=\hsize]{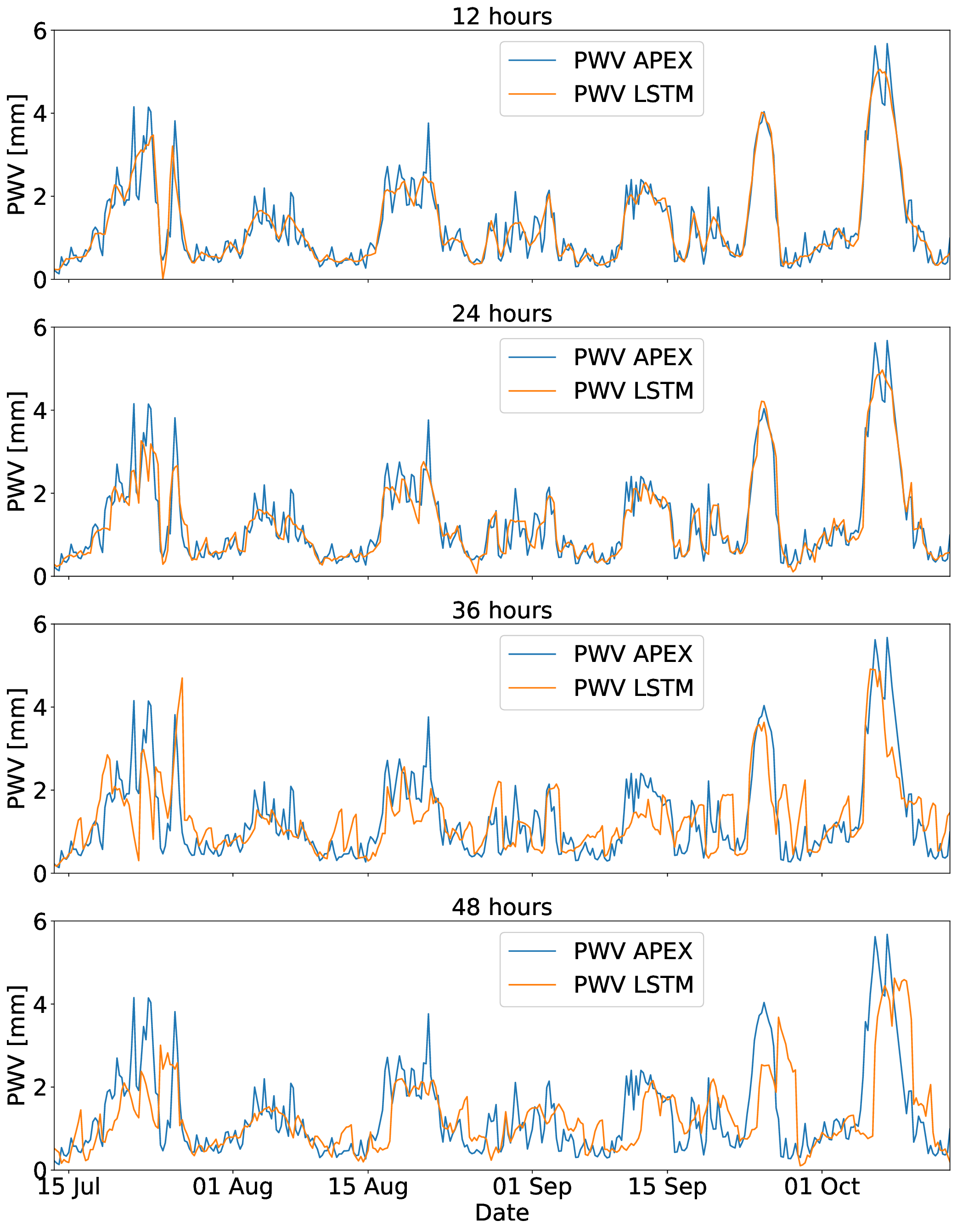}
      \caption{Comparison of PWV APEX (blue) and PWV LSTM (orange) for the different forecasting horizons.}
         \label{scatter-actvspred}
   \end{figure}

Figure \ref{scatter-actvspred} illustrates the temporal comparison of the values predicted by the LSTM model (PWV LSTM) and the real observed values recorded by APEX (PWV APEX) across the four forecasting horizons. The plots visually confirm that the model effectively captures the temporal dynamics and amplitude variations of PWV over time, particularly for the short-term forecasts (12 and 24 h) where the predicted series closely tracks the observations. A greater divergence between the predicted and real PWV series is observed at longer horizons (36 and 48 h), consistent with the increasing error metrics ($r$, $R^2$, RMSE, and MAE) and the sharp rise in MAPE previously noted.

   \begin{figure}[h!]
   \centering
   \includegraphics[width=\hsize]{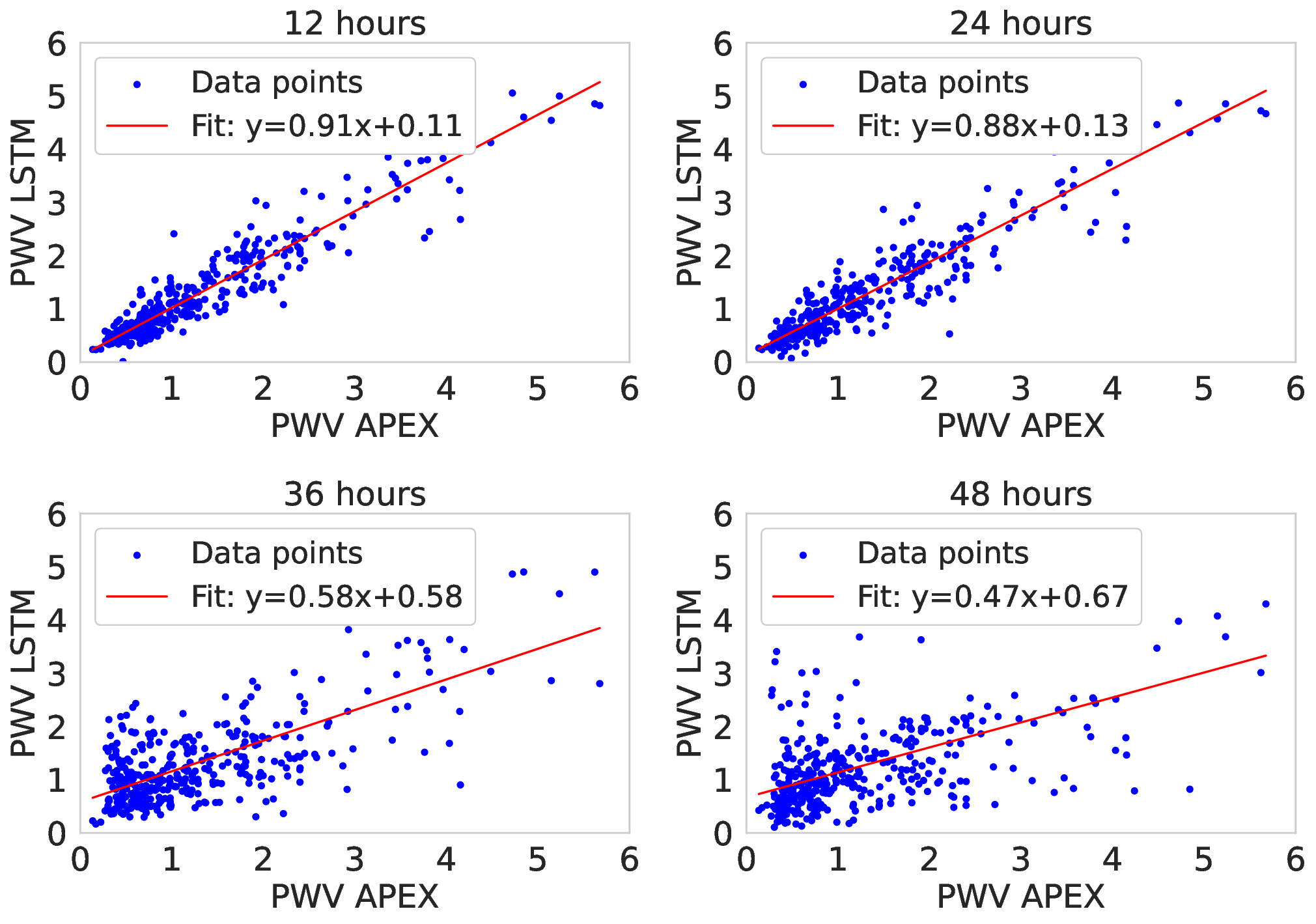}
      \caption{Scatter plots for PWV APEX and PWV LSTM at the four different forecasting horizons, including the linear regression fit.}
         \label{rmse}
   \end{figure}

Figure \ref{rmse} presents scatter plots of these predicted versus real PWV values for each forecasting horizon, including the regression line. The tight clustering of data points along the regression line for the 12 and 24\,h horizons reinforces the strong correlation and higher accuracy metrics reported. As the forecasting horizon increases to 36 and 48\,h, the scatter increases significantly, and the values of $r$ and $R^2$ drop substantially, indicating a weaker linear relationship and reduced predictive power, consistent with the quantitative metrics.

\begin{figure}[h!]
   \centering
   \includegraphics[width=1\hsize]{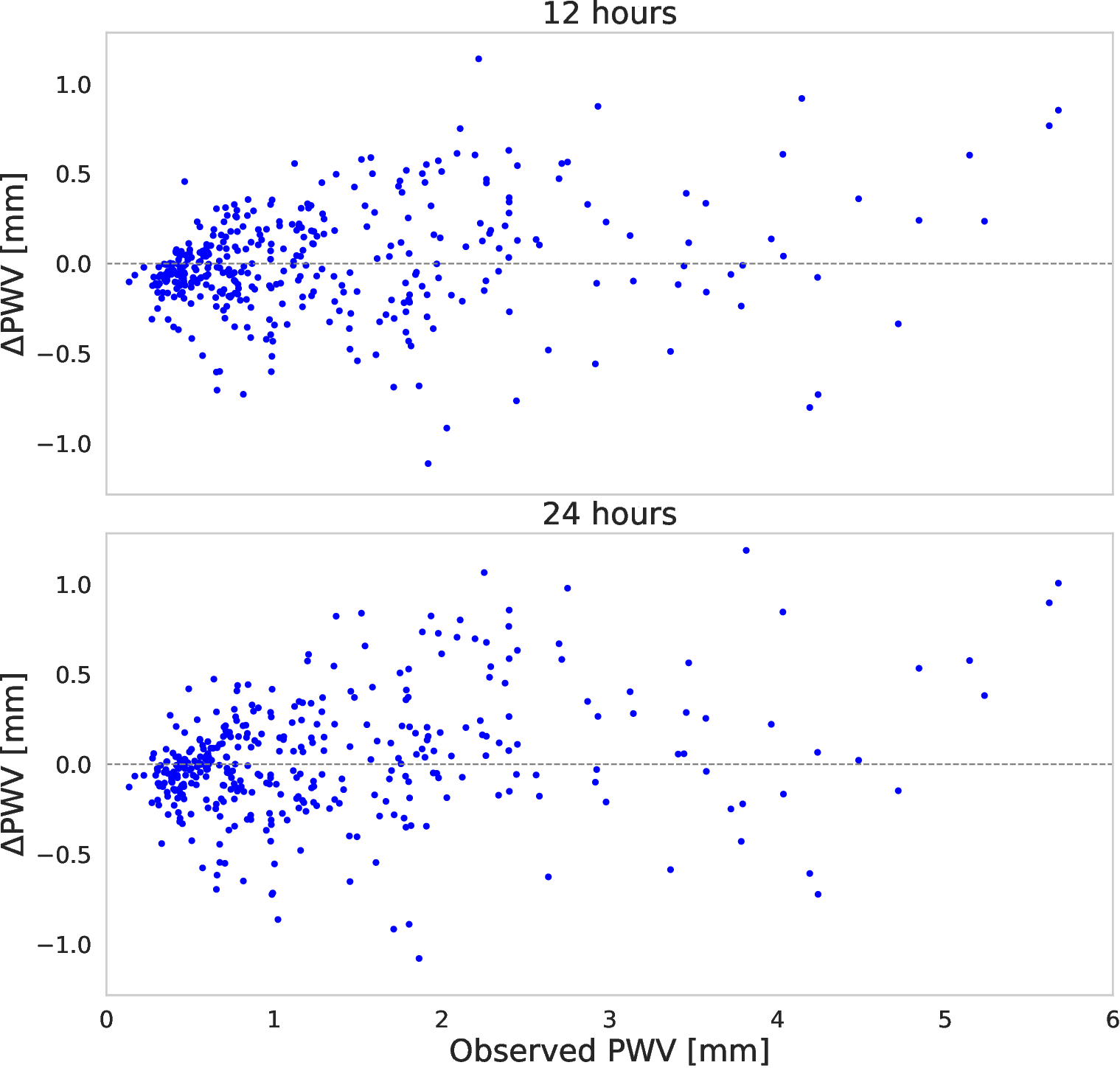}
      \caption{Scatter plots of $\Delta$PWV (observed minus predicted) against observed PWVs for 12 h (top) and 24 h (bottom) forecasting horizons. The distribution remains centered near zero, with increasing dispersion at higher PWV values.}
      \label{dpwv}
   \end{figure}

To analyze the distribution of prediction errors at different PWV levels, Figure~\ref{dpwv} shows the differences between observed and forecasted PWV ($\Delta$PWV = PWV$_{obs}$ – PWV$_{pred}$) as a function of the observed PWV values for the 12 and 24\,h horizons. The plots indicate that the errors remain centered around zero, with higher dispersion at larger PWV values.

To determine the performance and errors at different PWV values, we stratified RMSE values into three categories. For the 12\,h horizon, RMSE was 0.20\,mm under very dry conditions (PWV $<$ 1\,mm), 0.34\,mm for intermediate conditions (1–2\,mm), and 0.53\,mm for PWV $>$ 2\,mm. For the 24\,h horizon, errors increased slightly, with RMSE values of 0.22\,mm, 0.39\,mm, and 0.62\,mm for the same categories, respectively. These results confirm that the model performs best under dry-to-moderate PWV conditions, which are the most relevant for astronomical observations, while prediction errors grow as atmospheric humidity increases.

\subsection{Comparison of LSTM and GFS at 12 and 24\,h}
\begin{figure}[h!]
   \centering
   \includegraphics[width=\hsize]{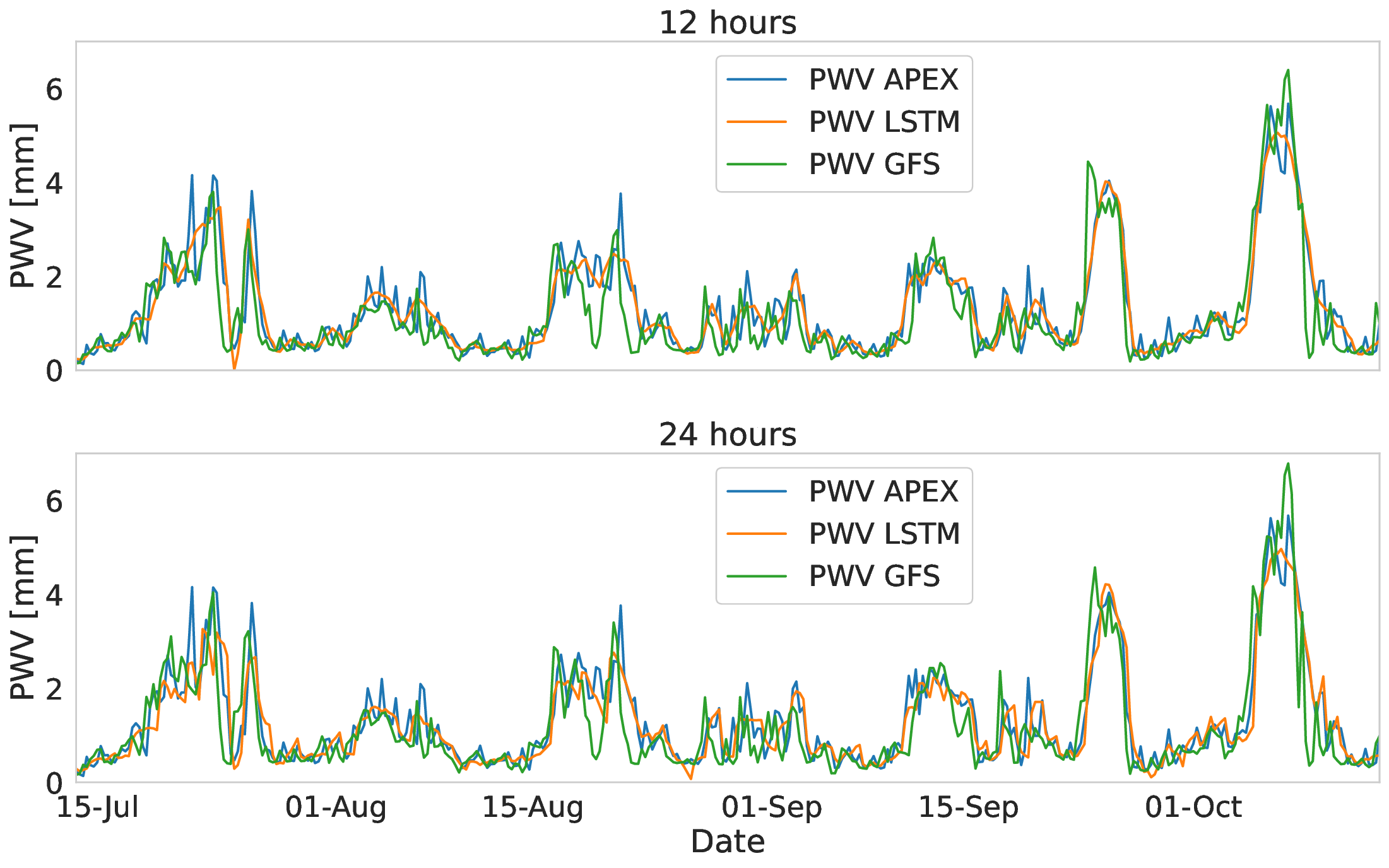}
      \caption{Comparison of PWV LSTM (orange), PWV GFS (green), and PWV APEX (blue) for 12 and 24\,h horizons.}
         \label{comp}
   \end{figure}

\begin{figure}[h!]
   \centering
   \includegraphics[width=\hsize]{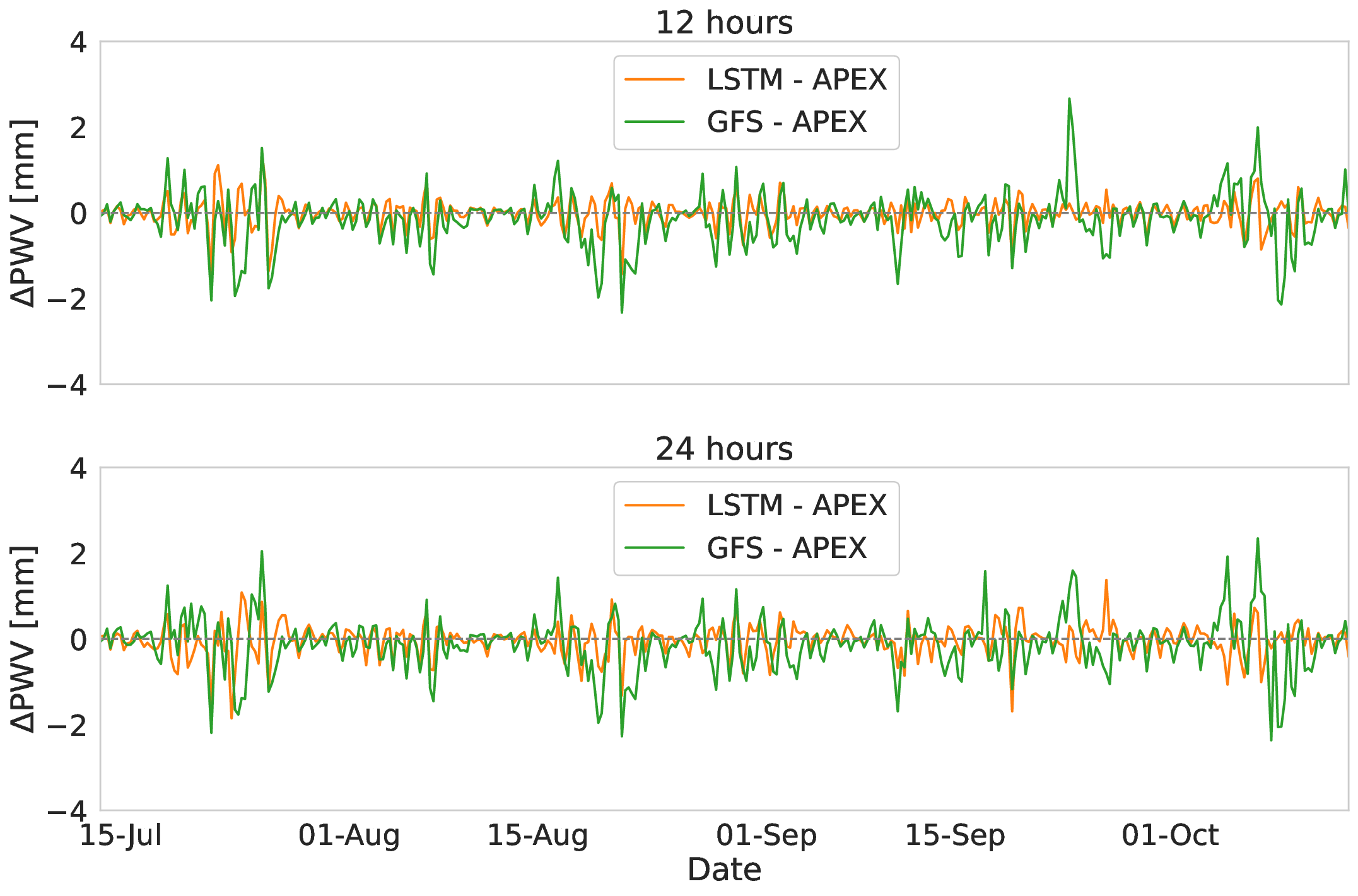}
      \caption{Comparisons of PWV LSTM (orange) and PWV GFS (green) with PWV APEX for 12 h and 24 h horizons.}
      \label{fig:diff_lstm_gfs_apex}
   \end{figure}

We compared our results with the GFS prediction model for the 12 and 24\,h forecasting horizons, which represent the most robust outcomes of the LSTM model. The GFS data (PWV GFS) are the predictions obtained at APEX with a total of 5-day forecasting horizon in steps of 6-hour resolution. We analyzed the LSTM and GFS prediction results in the same period of time, same time-resolution, and same forecasting horizons for a direct comparison of these models. Considering the 12 and 24\,h forecasting horizons to match with our LSTM predictions, Figure \ref{comp} compares both prediction models with PWV APEX at each forecasting horizon. The differences between the LSTM and GFS models with real values (PWV LSTM – PWV APEX and PWV GFS – PWV APEX) are shown in Figure \ref{fig:diff_lstm_gfs_apex}.

Table \ref{table:lstm_gfs} shows that the LSTM model is able to reduce RMSE in short-term forecasts (12 and 24\,h), achieving a $RMSE_{LSTM}$ of 0.328 and 0.376\,mm, respectively. These are notably lower than $RMSE_{GFS}$ of 0.625 and 0.636\,mm, significantly reducing the error by \string~50\%. Analyzing MAPE values, we find the overall error for LSTM to be within \string~22\% while for GFS are within \string~36\%. 

\begin{table}[h!]
\caption{Comparison of the RMSE (in mm) and MAPE (in \%) from the LSTM and GFS predictions with PWV APEX.}
\label{table:lstm_gfs}
\centering
\begin{tabular}{ccccc}
\hline\hline

Time & $RMSE_{LSTM}$ & $RMSE_{GFS}$ & $MAPE_{LSTM}$ & $MAPE_{GFS}$ \\

\hline
12 h & 0.328 & 0.625 & 21.1 & 35.6 \\
24 h & 0.376 & 0.636 & 23.2 & 36.2 \\
36 h & 0.768 & 0.662 & 62.8 & 36.8 \\
48 h & 0.916 & 0.644 & 65.9 & 36.0 \\
\hline
\end{tabular}
\end{table}

However, for 36 and 48\,h horizons, $RMSE_{LSTM}$ increases to 0.768 and 0.916\,mm, surpassing the GFS model with $RMSE_{GFS}$ of 0.662 and 0.644\,mm, respectively. At high PWV values, the differences between LSTM with APEX become greater. To enable an analysis in the same PWV range (0.4-1.2\,mm) as the GFS data reported in \citet{marin_estimating_2015}, we compared LSTM and GFS predictions in Table \ref{table:lstm_gfs_range}.

\begin{table}[h!]
\caption{Comparison of the different RMSE (in mm) and MAPE (\%) values of the LSTM and GFS predictions in the 0.4-1.2\,mm range.}
\label{table:lstm_gfs_range}
\centering
\begin{tabular}{ccccc}
\hline\hline

Time & $RMSE_{LSTM}$ & $RMSE_{GFS}$ & $MAPE_{LSTM}$ & $MAPE_{GFS}$\\

\hline
12 h & 0.235 & 0.356 & 23.2 & 35.9 \\
24 h & 0.245 & 0.385 & 25.1 & 36.3 \\
36 h & 0.631 & 0.393 & 67.9 & 36.1 \\
48 h & 0.595 & 0.362 & 63.3 & 34.4 \\
\hline
\end{tabular}
\end{table}

In the 0.4-1.2\,mm range, for the 12\,h forecasting horizon, the RMSE for LSTM presents a reduction of \string~34\% compared with GFS and for 24\,h, the reduction is \string~36\%. The RMSE for LSTM increases to \string~60\% for longer forecasting horizons at 36 and 48\,h. MAPE values at the 12 and 24 h forecasting horizons are \string~23 and 25\% for LSTM, while GFS shows similar percentage values of \string~35\% in all four forecasting horizons. These results reinforce the improved performance of the LSTM predictions at different PWV levels compared with traditional GFS predictions at 12 and 24\,h forecasting horizons. Larger discrepancies for LSTM are observed at longer forecasting horizons and the strategies proposed to enhance our model are discussed next.

\subsection{Discrepancies at longer forecast horizons (36 and 48\,h)}
\label{subsec:longer_horizons}

The comparison between predicted and observed PWVs at the 36 and 48\,h horizons consistently shows a marked degradation in predictive accuracy, with larger scatter, weaker correlations, and increasing errors relative to shorter horizons. This loss of performance can be attributed to several factors: (1) the cumulative propagation of errors across longer prediction windows, which is intrinsic to sequential models; (2) the increasing influence of small-scale, high-frequency atmospheric processes that are not captured well by the limited input variables; and (3) the reduced information content available to constrain forecasts beyond one day in such a highly variable environment. These limitations are consistent with known challenges in atmospheric forecasting.

\subsection{Strategies to improve short- and long-term forecasts}

Future upgrades and strategies are suggested to improve our PWV predictions in short (< 12 h) and long (> 36 h) timescales, focusing on several key areas: (1) expand the current dataset to include multiple years of historical observations to enhance model generalization and stability; (2) incorporate additional meteorological variables that are correlated with PWV, including historical meteorological reanalysis data, satellite data, solar radiation, cloudiness, air pressure, and others that are available in the Chajnantor area; (3) explore shorter time-resolution datasets and forecast horizons, for example 10 minutes or 1 hour, and larger forecasting horizons up to 1 week or more; (4) test alternative deep learning architectures such as attention-based models andtTemporal convolutional networks; and (5) assess real-time operational deployment potential in coordination with observatories scheduling systems. These strategies highlight possible paths to strengthen the predictive capacity of the model. Considering these upgrades and strategies in our model, a future plan is to produce real-time predictions at different forecasting horizons available on a public web page.

\section{Conclusions}
This study presents a first investigation of the application of deep learning techniques (specifically LSTM networks) for preliminary predictions of PWV in the Chajnantor area. Using radiometric and meteorological data collected locally for more than 1 year, a site-specific LSTM-based forecasting model was developed and evaluated.

The LSTM model demonstrated higher predictive accuracy for 12- and 24-hour forecasting horizons: it improved the traditional GFS method used by APEX in all $r$, $R^{2}$, RMSE, MAE, and MAPE metrics, reducing the prediction errors by \string~50\%. However, the LSTM model performance declined at the 36- and 48-hour horizon, with an increase in error variability and relative error. This degradation, while expected in time series forecasting, suggests that further improvements are needed before the model can reliably support longer-term astronomical observing planning. We will  propose several strategies to upgrade and improve our model on shorter (< 12 h) and longer (> 36 h) timescale forecasting horizons in a future article.

These results highlight the potential of deep learning as a valuable tool for short-term operational PWV forecasting at high-altitude observatory sites. The ability to produce localized high-resolution forecasts tailored to specific atmospheric conditions is particularly advantageous in environments where global numerical weather models struggle to capture fine-scale moisture variability.

Although this study is site-specific, the proposed methodology can be generalized to other astronomical observatory locations. To do so, high-quality, high-resolution local PWV and meteorological data would be needed for model training. Furthermore, the input variables and preprocessing steps may need to be adapted to the specific atmospheric characteristics of each site. With sufficient data and careful retraining, the LSTM-based approach demonstrated here can be transferred to different geographic regions and used as a flexible forecasting tool for PWV predictions to support astronomical operations.

\begin{acknowledgements}
      We acknowledge support from FIC-R IA 40036152-0, Gobierno Regional del Bio-Bío. We acknowledge support from China Chile Joint Research Fund CCJRF1803 and CCJRF2101. SR acknowledges support from ANID FONDECYT Iniciación 11221231. RB and SR acknowledge support from ANID FONDECYT Regular 1251819. JC acknowledges support from ANID FONDECYT Regular 1240843. PG is supported by Chinese Academy of Sciences South America Center for Astronomy (CASSACA) Key Research Project E52H540201. RR acknowledges support from ANID BASAL FB210003 (CATA) and from Núcleo Milenio TITANs (NCN2023-002). We also acknowledge the availability of meteorological data, PWV, and GFS forecasts provided by the APEX telescope website and by the ALMA observatory.
\end{acknowledgements}

\bibliographystyle{aa}
\bibliography{bibtex}

@article{bustos_parque_2014,
  author  = {Bustos, R. and Rubio, M. and Ot{\'a}rola, A. and Nagar, N.},
  title   = {Parque Astron{\'o}mico de Atacama: An ideal site for millimeter, submillimeter, and mid-infrared astronomy},
  journal = {PASP},
  year    = {2014},
  volume  = {126},
  number  = {946},
  pages   = {1126--1132},
  doi     = {10.1086/679330},
  url     = {https://doi.org/10.1086/679330}
}

@article{castro-almazan_precipitable_2016,
	address = {Edinburgh, United Kingdom},
	title = {Precipitable {Water} {Vapour} at the {Canarian} {Observatories} ({Teide} and {Roque} de los {Muchachos}) from routine {GPS}},
	url = {http://proceedings.spiedigitallibrary.org/proceeding.aspx?doi=10.1117/12.2232646},
	doi = {10.1117/12.2232646}, 
	urldate = {2025-10-10},
	author = {Castro-Almazán, Julio A. and Muñoz-Tuñón, Casiana and García-Lorenzo, Begoña and Pérez-Jordán, Gabriel and Varela, Antonia M. and Romero, Ignacio},
    journal = {Proc. SPIE},
    volume = {9910},
	editor = {Peck, Alison B. and Seaman, Robert L. and Benn, Chris R.},
	month = jul,
	year = {2016},
	pages = {99100P},
}

@article{chen_global_2021,
  author  = {Chen, Biyan and Yu, Wenkun and Wang, Wei and Zhang, Zhetao and Dai, Wujiao},
  title   = {A global assessment of precipitable water vapor derived from {GNSS} zenith tropospheric delays with {ERA5}, {NCEP} {FNL}, and {NCEP} {GFS} products},
  journal = {Earth and Space Science},
  year    = {2021},
  volume  = {8},
  number  = {8},
  pages   = {e2021EA001796},
  doi     = {10.1029/2021EA001796},
  url     = {https://doi.org/10.1029/2021EA001796}
}

@article{cortes_twenty_2020,
  author  = {Cort{\'e}s, F. and Cort{\'e}s, K. and Reeves, R. and Bustos, R. and Radford, S.},
  title   = {Twenty years of precipitable water vapor measurements in the Chajnantor area},
  journal = {A\&A},
  year    = {2020},
  volume  = {640},
  pages   = {A126},
  doi     = {10.1051/0004-6361/202037784},
  url     = {https://doi.org/10.1051/0004-6361/202037784}
}

@article{giordano_atmospheric_2013,
  author  = {Giordano, C. and Vernin, J. and V{\'a}zquez Ram{\'\i}o, H. and Mu{\~n}oz-Tu{\~n}{\'o}n, C. and Varela, A. M. and Trinquet, H.},
  title   = {Atmospheric and seeing forecast: WRF model validation with in situ measurements at ORM},
  journal = {MNRAS},
  year    = {2013},
  volume  = {430},
  number  = {4},
  pages   = {3102--3111},
  doi     = {10.1093/mnras/stt117},
  url     = {https://doi.org/10.1093/mnras/stt117}
}

@article{gusten_atacama_2006,
  author  = {G{\"u}sten, R. and Nyman, L. {\AA}. and Schilke, P. and Menten, K. and Cesarsky, C. and Booth, R.},
  title   = {The Atacama Pathfinder EXperiment (APEX)---a new submillimeter facility for southern skies},
  journal = {A\&A},
  year    = {2006},
  volume  = {454},
  number  = {2},
  pages   = {L13--L16},
  doi     = {10.1051/0004-6361:20065420},
  url     = {https://doi.org/10.1051/0004-6361:20065420}
}

@inproceedings{haputhanthri_short-term_2021,
  author    = {Haputhanthri, Dilantha and Wijayasiri, Adeesha},
  title     = {Short-term traffic forecasting using {LSTM}-based deep learning models},
  booktitle = {Moratuwa Engineering Research Conference (MERCon)},
  year      = {2021},
  pages     = {602--607},
  publisher = {IEEE},
  doi       = {10.1109/MERCon52712.2021.9525670},
  url       = {https://doi.org/10.1109/MERCon52712.2021.9525670}
}

@article{hochreiter_long_1997,
  author  = {Hochreiter, Sepp and Schmidhuber, J{\"u}rgen},
  title   = {Long short-term memory},
  journal = {Neural Comput.},
  year    = {1997},
  volume  = {9},
  number  = {8},
  pages   = {1735--1780},
  doi     = {10.1162/neco.1997.9.8.1735},
  url     = {https://doi.org/10.1162/neco.1997.9.8.1735}
}

@article{hou_machine_2023,
  author  = {Hou, X. and Hu, Y. and Du, F. and Ashley, M. C. B. and Pei, C. and Shang, Z. and Ma, B. and Wang, E. and Huang, K.},
  title   = {Machine learning-based seeing estimation and prediction using multi-layer meteorological data at Dome A, Antarctica},
  journal = {Astron. Comput.},
  year    = {2023},
  volume  = {43},
  pages   = {100710},
  doi     = {10.1016/j.ascom.2023.100710},
  url     = {https://doi.org/10.1016/j.ascom.2023.100710}
}

@article{jain_forecasting_2020,
  author    = {Jain, Mayank and Manandhar, Shilpa and Lee, Yee Hui and Winkler, Stefan and Dev, Soumyabrata},
  title     = {Forecasting precipitable water vapor using LSTMs},
  booktitle = {2020 IEEE USNC-CNC-URSI North American Radio Science Meeting (Joint with AP-S Symposium)},
  year      = {2020},
  journal = {Proc. IEEE AP-S and USNC-URSI Meeting},
  pages     = {147--148},
  publisher = {IEEE},
  doi       = {10.23919/USNC/URSI49741.2020.9321614},
  url       = {https://doi.org/10.23919/USNC/URSI49741.2020.9321614}
}

@article{li_investigating_2023,
  author  = {Li, Haobo and Choy, Suelynn and Wang, Xiaoming and Liang, Hong and Purwar, Smrati and Zhang, Kefei},
  title   = {Investigating the Optimal Spatial Resolution for Assimilating GNSS PWV into an NWP System to Improve the Accuracy of Humidity Field},
  journal = {IEEE J-STARS},
  year    = {2023},
  volume  = {16},
  pages   = {6876-6888},
  doi     = {10.1109/JSTARS.2023.3298489},
  url     = {https://ieeexplore.ieee.org/document/10192279/},
  issn    = {1939-1404}
}

@inproceedings{li_research_2024,
  author    = {Li, Yanchuan and Hu, Tong and Qi, Ke and Yan, Xundong and Li, Chunxian and Wang, Bo},
  title     = {Research on Real Time Single-Station GNSS PWV Monitoring Method and Experimental Verification at Sea},
  booktitle={14th International Symposium on Antennas, Propagation and EM Theory (ISAPE)},
  year      = {2024},
  publisher = {IEEE},
  pages     = {1--4},
  isbn      = {979-8-3503-5312-9},
  doi       = {10.1109/ISAPE62431.2024.10841001},
  url       = {https://ieeexplore.ieee.org/document/10841001/}
}

@article{marin_estimating_2015,
  author  = {Mar{\'\i}n, Julio C. and Pozo, Diana and Cur{\'e}, Michel},
  title   = {Estimating and forecasting the precipitable water vapor from {GOES} satellite data at high-altitude sites},
  journal = {A\&A},
  year    = {2015},
  volume  = {573},
  pages   = {A41},
  doi     = {10.1051/0004-6361/201424460},
  url     = {https://doi.org/10.1051/0004-6361/201424460}
}

@article{otarola_precipitable_2019,
  author  = {Ot{\'a}rola, A. and De Breuck, C. and Travouillon, T. and Matsushita, S. and Nyman, L.-{\AA}. and Wootten, A. and Radford, S. J. E. and Sarazin, M. and Kerber, F. and P{\'e}rez-Beaupuits, J. P.},
  title   = {Precipitable water vapor, temperature, and wind statistics at sites suitable for mm and submm wavelength astronomy in northern Chile},
  journal = {PASP},
  year    = {2019},
  volume  = {131},
  pages   = {045001},
  doi     = {10.1088/1538-3873/aafb78},
  url     = {https://doi.org/10.1088/1538-3873/aafb78}
}

@article{otarola_thirty_2010,
  author  = {Ot{\'a}rola, A. and Travouillon, T. and Sch{\"o}ck, M. and Els, S. and Riddle, R. and Skidmore, W. and Dahl, R. and Naylor, D. and Querel, R.},
  title   = {Thirty Meter Telescope site testing {X}: Precipitable water vapor},
  journal = {PASP},
  year    = {2010},
  volume  = {122},
  number  = {890},
  pages   = {470--484},
  doi     = {10.1086/651582},
  url     = {https://doi.org/10.1086/651582}
}

@article{pozo_validation_2016,
  author  = {Pozo, Diana and Mar{\'\i}n, J. C. and Illanes, L. and Cur{\'e}, M. and Rabanus, D.},
  title   = {Validation of WRF forecasts for the Chajnantor region},
  journal = {MNRAS},
  year    = {2016},
  volume  = {459},
  number  = {1},
  pages   = {419--426},
  doi     = {10.1093/mnras/stw600},
  url     = {https://doi.org/10.1093/mnras/stw600}
}

@article{perez-jordan_precipitable_2018,
	title = {Precipitable water vapour forecasting: a tool for optimizing {IR} observations at {Roque} de los {Muchachos} {Observatory}},
	volume = {477},
	copyright = {http://academic.oup.com/journals/pages/about\_us/legal/notices},
	issn = {0035-8711, 1365-2966},
	shorttitle = {Precipitable water vapour forecasting},
	url = {https://academic.oup.com/mnras/article/477/4/5477/4972781},
	doi = {10.1093/mnras/sty943},
	language = {en},
	number = {4},
	urldate = {2025-10-10},
	journal = {MNRAS},
	author = {Pérez-Jordán, wG and Castro-Almazán, J A and Muñoz-Tuñón, C},
	month = jul,
	year = {2018},
	pages = {5477--5485},
}

@article{radford_observing_2011,
  author        = {Radford, Simon J. E.},
  title         = {Observing conditions for submillimeter astronomy},
  journal       = {RevMexAA},
  year          = {2011},
  volume        = {41},
  pages         = {87--90},
  eprint        = {1107.5633},
  archivePrefix = {arXiv},
  primaryClass  = {astro-ph.IM},
  doi           = {10.48550/arXiv.1107.5633},
  url           = {https://arxiv.org/abs/1107.5633}
}

@article{radford_submillimeter_2016,
  author  = {Radford, Simon J. E. and Peterson, Jeffery B.},
  title   = {Submillimeter atmospheric transparency at Maunakea, at the South Pole, and at Chajnantor},
  journal = {PASP},
  year    = {2016},
  volume  = {128},
  pages   = {075001},
  doi     = {10.1088/1538-3873/128/965/075001},
  url     = {https://doi.org/10.1088/1538-3873/128/965/075001}
}

@article{sleem_new_2024,
  author  = {Sleem, Ragab Elhady and Abdelfatah, Mohamed Amin and Mousa, Ashraf El-Kutb and El-Fiky, Gamal Saber},
  title   = {A new Egyptian Grid Weighted Mean Temperature (EGWMT) model using hourly ERA5 reanalysis data in GNSS PWV retrieval},
  journal = {Sci Rep},
  year    = {2024},
  volume  = {14},
  number  = {1},
  pages   = {14608},
  doi     = {10.1038/s41598-024-64132-2},
  url     = {https://www.nature.com/articles/s41598-024-64132-2},
  issn    = {2045-2322}
}

@article{turchi_forecasting_2018,
  author  = {Turchi, Alessio and Masciadri, Elena and Kerber, Florian and Martelloni, Gianluca},
  title   = {Forecasting water vapour above the sites of {ESO}'s Very Large Telescope (VLT) and the Large Binocular Telescope (LBT)},
  journal = {MNRAS},
  year    = {2019},
  volume  = {482},
  pages   = {206--},
  doi     = {10.1093/mnras/sty2668},
  url     = {https://doi.org/10.1093/mnras/sty2668}
}

@article{valeria_satellite-based_2024,
  author  = {Valeria, L. and Mart{\'\i}nez-Ledesma, M. and Reeves, R.},
  title   = {Satellite-based atmospheric characterization for sites of interest in millimeter and sub-millimeter astronomy},
  journal = {A\&A},
  year    = {2024},
  volume  = {684},
  pages   = {A186},
  doi     = {10.1051/0004-6361/202347773},
  url     = {https://doi.org/10.1051/0004-6361/202347773}
}

@article{vaquero-martinez_review_2021,
	title = {Review on the {Role} of {GNSS} {Meteorology} in {Monitoring} {Water} {Vapor} for {Atmospheric} {Physics}},
	volume = {13},
	copyright = {https://creativecommons.org/licenses/by/4.0/},
	issn = {2072-4292},
	url = {https://www.mdpi.com/2072-4292/13/12/2287},
	doi = {10.3390/rs13122287},
	language = {en},
	number = {12},
	urldate = {2025-10-10},
	journal = {Remote Sens.},
	author = {Vaquero-Martínez, Javier and Antón, Manuel},
	month = jun,
	year = {2021},
	pages = {2287},
}

@article{wu_high-precision_2021,
  author  = {Wu, Mingliang and Jin, Shuanggen and Li, Zhicai and Cao, Yunchang and Ping, Fan and Tang, Xu},
  title   = {High-Precision GNSS PWV and Its Variation Characteristics in China Based on Individual Station Meteorological Data},
  journal = {Remote Sens.},
  year    = {2021},
  volume  = {13},
  number  = {7},
  pages   = {1296},
  doi     = {10.3390/rs13071296},
  url     = {https://www.mdpi.com/2072-4292/13/7/1296},
  issn    = {2072-4292},
  month   = {Mar}
}

@article{xiao_prediction_2022,
  author  = {Xiao, Xingxing and Lv, Weicai and Han, Yuchen and Lu, Fukang and Liu, Jintao},
  title   = {Prediction of {CORS} water vapor values based on the {CEEMDAN} and {ARIMA}-{LSTM} combination model},
  journal = {Atmosphere},
  year    = {2022},
  volume  = {13},
  number  = {9},
  pages   = {1453},
  doi     = {10.3390/atmos13091453},
  url     = {https://doi.org/10.3390/atmos13091453}
}

@article{yan_research_2024,
  author  = {Yan, Xiangrong and Yang, Weifang and Li, Yuzhao and Su, Xiaoning and Zhang, Liming and Lu, Xiaomin},
  title   = {Research on {GNSS}-{PWV} retrieval and its application in rainfall forecasting based on deep learning},
  journal = {Athens J. Sci.},
  year    = {2024},
  volume  = {11},
  number  = {3},
  pages   = {165--180},
  doi     = {10.30958/ajs.11-3-2},
  url     = {https://doi.org/10.30958/ajs.11-3-2}
}

@article{yang_adaptability_2022,
  author  = {Yang, Mo and Wang, Jing},
  title   = {Adaptability of financial time series prediction based on {BiLSTM}},
  journal = {Procedia Comput. Sci.},
  year    = {2022},
  volume  = {199},
  pages   = {18--25},
  doi     = {10.1016/j.procs.2022.01.003},
  url     = {https://doi.org/10.1016/j.procs.2022.01.003}
}

\end{document}